\documentclass[
aps,
nofootinbib,
superscriptaddress,
tightenlines,
notitlepage,
twocolumn,
showpacs,
floatfix,
]{revtex4-1}
\usepackage{amssymb,amsmath,bm,tensor,braket}
\usepackage[colorlinks]{hyperref}

\usepackage{graphicx}
\usepackage{dcolumn}
\usepackage{bm}
\usepackage{xcolor}

\newcommand{\PlanckMass}{M_{\rm Pl}}

\newcommand{\Tabriz}{\affiliation{Faculty of Physics, University of Tabriz,
Tabriz 51666-16471, Iran}}

\begin{document}

\preprint{APS/123-QED}

\title{Cosmology in Brans-Dicke-de Rham-Gabadadze-Tolley massive gravity}

\author{Sobhan Kazempour}\email{s.kazempour@tabrizu.ac.ir}\Tabriz
\author{Amin Rezaei Akbarieh}\email{am.rezaei@tabrizu.ac.ir}\Tabriz

\date{\today}

\begin{abstract}
We introduce the Brans-Dick de Rham-Gabadadze-Tolley massive gravity theory which is the new extension of nonlinear massive gravity. We demonstrate a detailed study of the cosmological properties of this theory of gravity, and we show the transformation of the Jordan frame to the
Einstein frame. We obtain the cosmological background equations and show the analyses of self-accelerating solutions for explaining the accelerated expansion of the Universe. In the following, we analyze the background perturbations, which consist of tensor, vector, and scalar perturbations within the framework of the new extension of the dRGT massive gravity in the Friedman-Lema\^itre-Robertson-Walker cosmology.
\end{abstract}


\maketitle


\section{\label{sec:intro}Introduction}

The majority of observational research shows that one of the greatest unsolved puzzles in cosmology is the late-time accelerated expansion of the Universe \cite{SupernovaSearchTeam:1998fmf,SNLS:2011cra,WMAP:2003xez,WMAP:2003elm,Caldwell:2003hz,Koivisto:2005mm,SDSS:2005xqv,SDSS:2009ocz}.
We know that in the framework of general relativity the accelerated expansion of the Universe is related to the unknown form of energy which is called dark energy \cite{Copeland:2006wr,Carroll:2003st,Bamba:2012cp,Kim:2005at}. While one of the options is the cosmological constant which was introduced by Albert Einstein, there is a big disagreement between the observed and the theoretical value of vacuum energy \cite{Weinberg:1988cp,Peebles:2002gy}.

In the modified gravity theory, there are several attempts for explaining the accelerated expansion of the Universe at late times. It is strongly believed that one of the valuable modifications of general relativity is the massive gravity theory, in this theory the gravity is propagated by a spin-2 nonzero graviton mass \cite{deRham:2010ik,deRham:2010kj,Hinterbichler:2011tt,deRham:2014zqa,Hassan:2011hr,Hassan:2011zd}.

Over the years, the massive gravity theory has spent phenomenological and cosmological ups and downs. After that Fierz and Pauli introduced the first linear ghost-free action in 1939 \cite{Fierz:1939ix}, it was demonstrated that this theory does not reduce to general relativity in the limit of zero graviton mass, i.e., van Dam-Veltman-Zakharov discontinuity \cite{vanDam:1970vg,Zakharov:1970cc}. It is interesting to mention that the vDVZ discontinuity has been looked at from different angels \cite{Porrati:2002cp,Jaccard:2013gla,Modesto:2013jea,Myung:2017zsa}. In order to avoid the vDVZ discontinuity, Vainshtein proposed the nonlinear Fierz-Pauli action instead of linear \cite{Vainshtein:1972sx}. While Boulware and Deser claimed the nonlinear Fierz-Pauli action has a ghost \cite{Boulware:1972yco}, in 2010 de Rham, Gabadadze, and Tolley (dRGT) exhibited the ghost-free nonlinear massive gravity \cite{deRham:2010ik,deRham:2010kj}.
On the one hand, the dRGT massive gravity can explain the accelerated expansion of the Universe without dark energy, and it is only valid for an open FLRW solution. On the other hand, there are not any stable solutions for homogeneous and isotropic Universe \cite{DeFelice:2012mx}. Furthermore, because of a strong coupling problem and a nonlinear ghost instability the scalar and vector perturbations would be vanished in this theory \cite{Gumrukcuoglu:2011zh}.

A huge number of scientists have been motivated to find a satisfactory massive gravity theory using changing the background or the original theory in different ways. In fact, breaking either homogeneity or isotropy of the background is one of the ways \cite{DAmico:2011eto,Gumrukcuoglu:2012aa,DeFelice:2013awa}. Moreover, other ways are, adding the new degrees of freedom or changing the effective parameters of the theory \cite{DAmico:2012hia,Huang:2012pe,Comelli:2013txa,Langlois:2014jba,deRham:2014gla,deRham:2014naa,EmirGumrukcuoglu:2014uog,DeFelice:2016tiu,Lin:2017oow,Kenna-Allison:2018izo,Kenna-Allison:2019tbu,Gumrukcuoglu:2020utx,Akbarieh:2021vhv,Aslmarand:2021qwn,Akbarieh:2022ovn}.
In this present work, we perform a detailed analysis of Brans-Dick dRGT massive gravity which is the new extension of nonlinear dRGT massive gravity theory. Thus, we try to demonstrate that the accelerated expansion of the Universe can be explained in the FLRW cosmology in the framework of this theory, and the perturbations analysis is free of instability. 

The Brans-Dicke theory is one of the many scalar-tensor alternative theories to the standard Einstein general relativity.
Note that the Brans-Dicke gravity theory introduces an additional long-range scalar field $\sigma$ besides the metric tensor $g_{\mu\nu}$ of spacetime. This theory can be considered a viable alternative to general relativity, one which compatible with Mach’s principle. It is interesting to point out that the scalar field does not exert any direct influence on matter, its role is that of participating in the field equations that determine the geometry of spacetime. 
It is noticeable that the Brans-Dicke theory describes gravitation in terms of a scalar field \cite{Brans:1961sx}. While the singularity problem remains in Brans-Dicke theory, all the available observational and experimental tests are being passed \cite{will2018theory}. As naturally can be seen in the string theory, the scalar field provides the local dynamical degree of freedom to the Brans-Dick theory \cite{Sa:1996ty,CamposDias:2003tv}.
\\
According to the observations, the original Brans-Dicke theory can not explain the cosmic acceleration \cite{Brans:1961sx,SupernovaSearchTeam:1998fmf,Dalal:2000xw,WMAP:2003elm}. 
To obtain an accelerating universe, it could be possible to modify this theory in different ways. Several extended Brans-Dicke theories have been investigated elaborately\cite{Xu:2008sn,Banerjee:2000mj,DeFelice:2010jn,Lu:2012zp,Roy:2017mnz,Hrycyna:2013yia,Ozer:2017oik,Freitas:2011st,Zhang:2017sym,Tripathy:2014spa,Papagiannopoulos:2016dqw,Sharif:2016glf,Kim:2004wj,Lee:2022cyh}.

The goals of this paper are, to find the self-accelerating solution for the late-time accelerated expansion of the Universe and to show the perturbations analysis. We introduce the Brans-Dick dRGT massive gravity in the Jordan frame and we transform it into the Einstein frame by maintaining the invariance of physical laws under units transformations.
\\
The paper is organized as follows. In Sec. \ref{sec:1}, we introduce the Brans-Dick dRGT massive gravity theory, and we show the transformation of the Jordan frame to the Einstein frame. Moreover, we obtain background equations and self-accelerating solutions. In Sec. \ref{sec:5}, we analyze the cosmological perturbations to demonstrate the tensor, vector, and scalar perturbations. Finally, in Sec. \ref{sec:6}, we conclude with a discussion. 
\\
Here, we define the $\PlanckMass^{2}\equiv 8\pi G =1$ and, $G$ is Newton’s constant.
We use units in which the speed of light and the reduced Planck constant assume the
value unity. We will assume natural units ($c = \hslash = 1$).

\section{\label{sec:1} Brans-Dick dRGT Massive Gravity}

In this section, we review the Brans-Dicke dRGT massive gravity theory and show the details of the conformal transformation. Also, we discuss the evolution of a cosmological background for this theory.
We start with the Jordan frame of Brans-Dicke gravity which is extended by dRGT massive gravity. The action includes the Brans-Dicke-like field $\varphi$ which is a scalar field, the Ricci scalar $R$, the function $\omega(\varphi)$ is the Brans-Dicke coupling, a dynamical metric $g_{\mu\nu}$ and it's determinant $\sqrt{-g}$. Moreover, the last part of the action is related to the massive gravity theory which will be introduced in the following. The action is given by
\begin{eqnarray}\label{Action}
S&&=\frac{1}{2}\int d^{4} x \sqrt{-g}\Bigg\{\varphi R - \frac{\omega(\varphi)}{\varphi}\partial^{\mu}\varphi\partial_{\mu}\varphi +2 m_{g}^{2}U(\mathcal K)\Bigg\}.\nonumber\\
\end{eqnarray}
In this stage, we use the conformal transformation to the minimally coupled case for the Brans-Dicke field \cite{Dicke:1961gz}. We rescale the metric tensor as
\begin{eqnarray}
g_{\mu\nu}=\lambda^{-1}\tilde{g}_{\mu\nu} , \nonumber\\ \sqrt{-g}=\lambda^{-2}\sqrt{-\tilde{g}}.
\end{eqnarray}
Also, the other rescaling parameters are defined below
\begin{eqnarray}
\varphi = \lambda\tilde{\varphi},
\end{eqnarray}
\begin{eqnarray}
\omega(\varphi)=\omega(\tilde{\varphi}),
\end{eqnarray}
\begin{eqnarray}\label{U-MG}
U(k)= \lambda^{2}\tilde{U}(k).
\end{eqnarray}
The conformal transformation affects the lengths of time-like intervals and the norm of time-like vectors. However, it keeps the light coins unchanged \cite{wald1984}.
\\
Now, we introduce the conformal transformation of the Ricci scalar $R$ and the second part of the action as below \cite{Dicke:1961gz,Synge:1960},
\begin{eqnarray}\label{R}
R=\lambda(\tilde{R}+3\tilde{\Box}\ln{\lambda}-\frac{3}{2}\lambda^{-2}\partial_{\mu}\lambda\partial_{\nu}\lambda \tilde{g}^{\mu\nu}),
\end{eqnarray}
\begin{eqnarray}\label{phi}
\frac{1}{\varphi}\partial_{\mu}\varphi\partial^{\mu}\varphi =\frac{\lambda^{2}}{\tilde{\varphi}}\partial_{\mu}\tilde{\varphi}\partial^{\mu}\tilde{\varphi}+2\lambda\partial_{\mu}\lambda\partial^{\mu}\tilde{\varphi}+\tilde{\varphi}\partial_{\mu}\lambda\partial^{\lambda}\lambda,
\end{eqnarray}
where
\begin{eqnarray}
\tilde{\Box}\ln{\lambda}=\frac{1}{\sqrt{-\tilde{g}}}\big(\sqrt{-\tilde{g}}\tilde{g}^{\mu\nu}\lambda^{-1}\partial_{\mu}\lambda\big)_{,\nu},
\end{eqnarray}
here $\tilde{\Box}\equiv \tilde{g}^{\mu\nu}\tilde{\nabla}_{\mu}\tilde{\nabla}_{\nu}$ is d'Alembert's operator, and the $\tilde{\nabla}_{\mu}$ is the covariant derivative operator of the rescaled metric $\tilde{g}_{\mu\nu}$.
By substituting the Eq. (\ref{R}) and Eq. (\ref{phi}) in Eq. (\ref{Action}), we have
\begin{eqnarray}\label{Action2}
S=\frac{1}{2}\int d^{4} x\sqrt{-\tilde{g}} \Bigg\lbrace\tilde{\varphi}\tilde{R}+3\tilde{\varphi}\tilde{\Box}\ln{\lambda}
-(\omega(\tilde{\varphi}) +\frac{3}{2})\tilde{\varphi}\frac{1}{\lambda^{2}}\partial_{\mu}\lambda\partial^{\mu}\lambda\nonumber\\
-2\omega(\tilde{\varphi})\frac{\partial_{\mu}\lambda\partial^{\mu}\tilde{\varphi}
}{\lambda}-\frac{\omega(\tilde{\varphi})}{\tilde{\varphi}}\partial_{\mu}\tilde{\varphi}\partial^{\mu}\tilde{\varphi}+2m_{g}^{2}\tilde{U}(k)\Bigg\rbrace.\nonumber\\
\end{eqnarray}
Note that it can be possible we consider $\lambda$ as a function of $\varphi$, therefore $\tilde{\varphi}$ is a constant \cite{Dicke:1961gz}. 
\begin{equation}
\lambda =\frac{\varphi}{\tilde{\varphi}}.
\end{equation}
As the $\tilde{\varphi}$ is constant, after using the ordinary divergence $\sqrt{-\tilde{g}}\tilde{\Box}\ln{\lambda}$, we have
\begin{eqnarray}\label{Action3}
S=\frac{1}{2}\int d^{4} x\sqrt{-\tilde{g}} \Bigg\lbrace\tilde{R}-\frac{(\omega(\tilde{\varphi}) +\frac{3}{2})}{\lambda^{2}}\partial^{\mu}\lambda\partial_{\mu}\lambda +\frac{2 m_{g}^{2}\tilde{U}(k)}{\tilde{\varphi}}\Bigg\rbrace ,\nonumber\\
\end{eqnarray}
also, we redefine the $\lambda$, as below
\begin{eqnarray}
\lambda =e^{\sigma},\quad \partial_{\mu}\sigma=\frac{\partial_{\mu}\lambda}{\lambda}.
\end{eqnarray}
By substituting into the action Eq. (\ref{Action2}) and considering Eq. (\ref{U-MG}), we have
\begin{eqnarray}\label{Action4}
S=\frac{1}{2}\int d^{4} x\sqrt{-\tilde{g}} \Bigg\lbrace \tilde{R}-(\omega(\sigma) +\frac{3}{2})\partial^{\mu}\sigma\partial_{\mu}\sigma \nonumber\\+\frac{2 m_{g}^{2}}{\tilde{\varphi}}e^{-2\sigma}U(\mathcal K)\Bigg\rbrace .\nonumber\\
\end{eqnarray}
In this stage, as the $\tilde{\varphi}$ is a constant, we can consider it $\tilde{\varphi}=1$. Note that a tilde denotes quantities defined in the Einstein frame, in the following, we disregard the tilde ($\tilde{•}$) for simplifying our calculations, so we have the action in the Einstein frame
\begin{eqnarray}\label{Action5}
S=\frac{1}{2}\int d^{4} x\sqrt{-g} \Bigg\lbrace R-(\omega(\sigma) +\frac{3}{2})\partial^{\mu}\sigma\partial_{\mu}\sigma \nonumber\\+2 m_{g}^{2}e^{-2\sigma}U(\mathcal K)\Bigg\rbrace .\nonumber\\
\end{eqnarray}
In the following, we introduce the $U({\cal K})$. It is obvious that the mass of graviton comes up with the potential $U$ which consists of three parts \cite{deRham:2010kj}.
\begin{equation}\label{Upotential1}
U(\mathcal{K})=U_{2}+\alpha_{3}U_{3}+\alpha_{4}U_{4},
\end{equation}
where $\alpha_3$ and $\alpha_4$ are dimensionless free parameters of the
theory. $U_i$ ($i=2,3,4$) is given by,
\begin{eqnarray}\label{Upotential2}
U_{2}&=&\frac{1}{2}\big([\mathcal{K}]^{2}-[\mathcal{K}^{2}]\big),
\nonumber\\
U_{3}&=&\frac{1}{6}\big([\mathcal{K}]^{3}-3[\mathcal{K}][\mathcal{K}^{2}]+2[\mathcal{K}^{3}]\big),
\nonumber\\
U_{4}&=&\frac{1}{24}\big([\mathcal{K}]^{4}-6[\mathcal{K}]^{2}[\mathcal{K}^{2}]+8[\mathcal{K}][\mathcal{K}^{3}]+3[\mathcal{K}^{2}]^2\nonumber\\&&-6[\mathcal{K}^{4}]\big),
\end{eqnarray}
where the quantity ``$[\cdot]$'' is interpreted as the trace of the tensor
inside brackets. It should be mentioned that the building block tensor
$\mathcal{K}$ is defined as
\begin{equation}\label{K}
\mathcal{K}^{\mu}_{\nu} = \delta^{\mu}_{\nu} -
\big(\sqrt{g^{-1}f}\big)_{~\nu}^{\mu},
\end{equation}
where $ f_{\alpha\nu}$ is the fiducial metric, which is defined through
\begin{equation}\label{7}
f_{\alpha\nu}=\partial_{\alpha}\phi^{c}\partial_{\nu}\phi^{d}\eta_{cd}.
\end{equation}
Here $g^{\mu\nu} $ is the physical metric, $\eta_{cd}$ is the Minkowski
metric with $c,d= 0,1,2,3$ and $\phi^{c}$ are the Stueckelberg fields which
are introduced to restore general covariance. 
According to our cosmological application purpose, we adopt the
Friedman-Lema\^itre-Robertson-Walker (FLRW) Universe. So, the general
expression of the corresponding dynamical and fiducial metrics are given as
follows,
\begin{align}
\label{DMetric}
g_{\mu\nu}&={\rm diag} \left[-N^{2},a^2,a^2,a^2 \right], \\
\label{FMetric}
f_{\mu\nu}&={\rm diag} \left[-\dot{f}(t)^{2},1,1,1 \right].
\end{align}
Note that we redefine the fiducial metric as $\tilde{f}=\lambda^{-1}f$. Thus, the relation of $\tilde{g}^{-1}\tilde{f}$ should be invariant.
Here, it is worth noting that $N$ is the lapse function of the
dynamical metric, and it is similar to a gauge function. Also, it is clear
that the scale factor is represented by $a$, and $\dot{a}$ is the
derivative with respect to time. Furthermore, the lapse function relates
the coordinate-time $dt$ and the proper-time $d\tau$ via $d\tau=Ndt$
\cite{Scheel:1994yr,Christodoulakis:2013xha}. Function $f(t)$ is the
Stueckelberg scalar function whereas $\phi^{0}=f(t)$ and
$\frac{\partial\phi^{0}}{\partial t}=\dot{f}(t)$ \cite{Arkani-Hamed:2002bjr}.
Therefore, the point-like Lagrangian of the Brans-Dicke dRGT massive gravity in FLRW
cosmology is given by
\begin{eqnarray}
\mathcal{L}=&&\Bigg\lbrace -\frac{3\dot{a}^{2}a}{N}+\frac{\big(2\omega(\sigma) +3\big)a^{3}\dot{\sigma}^{2}}{4N}\Bigg\rbrace \nonumber\\&& + \frac{m_{g}^{2} a (X-1)}{X^{2}}\Bigg\lbrace \bigg[3(X-2)-(X-4)(X-1)\alpha_{3}\nonumber\\&&-(X-1)^{2}\alpha_{4}\bigg]N+\dot{f}(t) a X\bigg[3-3(X-1)\alpha_{3}\nonumber\\&&+(X-1)^{2}\alpha_{4}\bigg]\Bigg\rbrace,\nonumber\\
\end{eqnarray}
where
\begin{equation}\label{XX}
X\equiv\frac{e^{\sigma}}{a}.
\end{equation}
In order to simplify expressions later, we define
\begin{equation}
H\equiv\frac{\dot{a}}{Na}.
\end{equation}

As we do not consider the matter stress tensor, however, there is in the realistic theory. It should be noted that in the Einstein frame both the Brans-Dicke scalar and the helicity-0 mode of the massive graviton would couple to a matter stress-tensor in the linearized approximation. The helicity-0 would be screened by the Vainshtein mechanism to avoid the 5th force, and it is completely similar to the quasi-dilation \cite{DAmico:2012hia}.

\subsection{Background Equations of Motion}\label{subsec4}

By considering the unitary gauge (i.e., $f(t)=t$), we obtain a constraint equation by varying with respect to $f$. It is worth pointing out that the gauge transformations eliminate the unphysical fields from the Lagrangian on the classical level \cite{Grosse-Knetter:1992tbp}. Thus, we achieve a constraint equation
\begin{eqnarray}\label{Cons}
\hspace{-0.5cm} \frac{\delta \mathcal{L}}{\delta f}= m_{g}^{2} \frac{d}{dt}\bigg[ && a^{2}\frac{(X-1)}{X}\nonumber\\
\hspace{-0.5cm} && \times[3-3(X-1)\alpha_{3}+(X-1)^{2}\alpha_{4}]\bigg]=0. \nonumber\\
\end{eqnarray}
In the following, the Friedman equation is derived by varying with respect to
the lapse function $N$,
\begin{eqnarray}\label{EqN}
\frac{1}{a^{3}}\frac{\delta\mathcal{L}}{\delta N}=&&3H^{2}-\frac{(2\omega(\sigma) +3)}{4}\big(H+\frac{\dot{X}}{NX}\big)^{2}\nonumber\\ && -m_{g}^{2}\frac{(X-1)}{X^{2}a^{2}}\bigg[-3(X-2)+(X-4)(X-1)\alpha_{3}\nonumber\\&&+(X-1)^{2}\alpha_{4}\bigg]=0.
\end{eqnarray}
Taking the variation of action Eq. (\ref{Action5}) with respect to the scalar field, the equation of motion corresponding to $\sigma$ is achieved as
\begin{widetext}
\begin{eqnarray}\label{EqSig}
\frac{1}{a^{3}N}\frac{\delta\mathcal{L}}{\delta \sigma}=&&\Bigg\lbrace \frac{m_{g}^{2}}{r^{2}X^{2}N^{2}}\bigg[ -2\big(6+4\alpha_{3}+\alpha_{4}\big)+(3+r)\big(3+3\alpha_{3}
+\alpha_{4}\big)X\nonumber\\&&-(3r+1)(\alpha_{3}+\alpha_{4})X^{3}+2r\alpha_{4}X^{4}
\bigg]-\frac{\bigg(N(6\omega(\sigma)+9)+\omega^{'}(\sigma)\bigg)H^{2}}{2N^{2}}\Bigg\rbrace =0, \nonumber\\
\end{eqnarray}
\end{widetext}
where
\begin{eqnarray}
r\equiv\frac{a}{N}.
\end{eqnarray}
The following equations can be achieved by using the notation in Eq. (\ref{XX})
\begin{equation}
\frac{\dot{\sigma}}{N}= H+\frac{\dot{X}}{NX}, \qquad \ddot{\sigma}=\frac{d}{dt}\Big(NH+\frac{\dot{X}}{X}\Big).
\end{equation}
Note that the Stueckelberg field $f$ introduces time reparametrization invariance. As a result, there is a Bianchi identity that relates the four equations of motion,
\begin{eqnarray}
\frac{\delta S}{\delta \sigma}\dot{\sigma}+\frac{\delta S}{\delta f}\dot{f}-N\frac{d}{dt}\frac{\delta S}{\delta N}+\dot{a}\frac{\delta S}{\delta a}=0.
\end{eqnarray}
Thus, one equation is redundant which is related to the varying with respect to the scale factor $a$, and can be eliminated.
Note that in the particular condition, all of the background equations and total Lagrangian reduce to those in Ref. \cite{DAmico:2012hia,Gumrukcuoglu:2013nza}

\subsection{Self-accelerating Background Solutions}\label{subsec5}

In this stage, we try to indicate the self-accelerating solutions elaborately.
After integrating the Stueckelberg constraint Eq. (\ref{Cons}) we have
\begin{eqnarray}\label{Self}
(1-\frac{1}{X})\bigg[3-3(X-1)\alpha_{3}+(X-1)^{2}\alpha_{4}\bigg] \propto a^{-2}. \nonumber\\
\end{eqnarray}
The constant solutions of $X$ lead to the effective energy density and behave similarly to a cosmological constant.
By considering an expanding universe, the right-hand side of that equation decrease as we have $a^{-2}$ in Eq.~(\ref{Self}).
After a long enough time, $X$ leads to a constant value, $X_{\rm
SA}$, which is a root of the left-hand side of Eq.~(\ref{Self}).
Here, we should pay attention that one obvious solution is $X=1$ which leads to a vanishing cosmological constant, and because of inconsistency it is unacceptable. So, this solution should
be discarded \cite{DAmico:2012hia}.
\begin{equation}
\big[3-3(X-1)\alpha_{3}+(X-1)^{2}\alpha_{4}\big]\bigg|_{X=X_{\rm SA}}=0.
\end{equation}
Thus, the two remaining solutions of Eq. (\ref{Self}) are
\begin{equation}\label{XSa}
X_{\rm SA}^{\pm}=\frac{3\alpha_{3}+2\alpha_{4}\pm\sqrt{9\alpha_{3}^{2}-12\alpha_{4}}}{2\alpha_{4}}.
\end{equation}
The Friedman equation (\ref{EqN}) could be written in a different form,
\begin{eqnarray}\label{EqFr}
\bigg(3-\frac{2\omega(\sigma) +3}{4}\bigg){H^{2}} = \Omega_{\mathcal{C}}, \qquad \Omega_{\mathcal{C}}=\frac{\Lambda_{\rm SA}^{\pm}}{a^{2}},
\end{eqnarray}
where
\begin{eqnarray}
\Lambda_{\rm SA}^{\pm}\equiv && m_{g}^{2}\frac{(X_{\rm SA}^{\pm}-1)}{X_{\rm SA}^{\pm ~ 2}}\Bigg[ -3X_{\rm SA}^{\pm}\nonumber\\&&
 +6+(X_{\rm SA}^{\pm}-4)(X_{\rm SA}^{\pm}-1)\alpha_{3}+(X_{\rm SA}^{\pm}-1)^{2}\alpha_{4}\Bigg].\nonumber\\
\end{eqnarray}
According to Eq. (\ref{XSa}), the above equation can be written as
\begin{widetext}
\begin{eqnarray}
\Lambda_{\rm SA}^{\pm}=\pm \frac{6 m_{g}^{2}\bigg(\pm 9\alpha_{3}^{4}+3\alpha_{3}^{3}\sqrt{9\alpha_{3}^{2}-12\alpha^{4}}\mp 18\alpha_{3}^{2}\alpha_{4}-4\alpha_{3}\sqrt{9\alpha_{3}^{2}-12\alpha_{4}}\pm 6\alpha_{4}^{2}\bigg)}{\alpha_{4}\bigg(\pm 3\alpha_{3}+\sqrt{9\alpha_{3}^{2}-12\alpha_{4}}\pm 2\alpha_{4}\bigg)^{2}}.
\end{eqnarray}
\end{widetext}
It should be noted that if we consider the $\omega(\sigma)$ as a constant, this condition imposes the curvature singularities. In other words, at finite values of $a$, the right-hand side of Eq. (\ref{EqFr}) goes to zero.
Note that when the $a$ is finite, the Hubble parameter, $\dot{a}$ and the scalar field are increasing. Thus, we have a real curvature singularity which is a big brake. In this condition, the universe reaches the finite scale factor and gets stuck. It can be found that similar types
of singularities \cite{Barrow:2004xh,Dabrowski:2007dn,Gregory:2007xy}.
In order to avoid the curvature singularity, we consider $\omega(\sigma)$ as an arbitrary function to remove the scale factor. Therefore, the curvature singularity is eliminated which means the self-accelerating solutions can be explained by an effective cosmological constant.

It is interesting to note that using Eq. (\ref{EqSig}), we calculate the $r_{\rm SA}$
\begin{widetext}
\begin{eqnarray}\label{rS}
r_{\rm SA}=&&\frac{1}{H^{2}X_{\rm SA}^{\pm ~ 2}\big(6\omega(\sigma)+ N\omega^{'}(\sigma)+9\big)}\Bigg\lbrace 3 m_{g}^{2}X_{\rm SA}^{\pm ~ 2}\big(\alpha_{3}X_{\rm SA}^{\pm}-\alpha_{3}-2\big)N\pm \bigg[m_{g}^{2}X_{\rm SA}^{\pm ~ 2}N\bigg(9 m_{g}^{2}X_{\rm SA}^{\pm ~ 2}(2+\alpha_{3}\nonumber\\&&-\alpha_{3}X_{\rm SA}^{\pm})^{2}N-2H^{2}\big(2(\alpha_{3}+3)+X_{\rm SA}^{\pm}(\alpha_{3}(X_{\rm SA}^{\pm}+3)-6(\alpha_{3}+2))\big)(6\omega(\sigma)+ N\omega^{'}(\sigma)+9)\bigg)\bigg]^{\frac{1}{2}}\Bigg\rbrace.
\end{eqnarray}
\end{widetext}
As there is not any strong coupling, this equation interprets the self-accelerating universe.
As a result, we have shown that this theory possesses self-accelerating solutions with an effective cosmological constant which is given by Eq. (\ref{EqFr}).
\\

\section{Perturbations Analysis}\label{sec:5}

The perturbations analysis is the essential tool for determining the stability of the solutions.
In fact, we would like to pay attention to the quadratic perturbations. 

In order to find the quadratic perturbations, the physical metric $g_{\mu\nu}$ could be expanded in terms of small fluctuations $\delta g_{\mu\nu}$ around a background solution $g_{\mu\nu}^{(0)}$.
\begin{equation}
g_{\mu\nu}=g_{\mu\nu}^{(0)}+\delta g_{\mu\nu}.
\end{equation}
Moreover, we can divide the metric perturbations into three parts,
namely scalar, vector, and tensor perturbations. So, we have
\begin{eqnarray}
\delta g_{00}=&&-2N^{2} \Phi, \nonumber\\
\delta g_{0i}=&&Na(B_{i}+\partial_{i}B), \nonumber\\
\delta g_{ij}=&&a^{2}\bigg[h_{ij}+\frac{1}{2}(\partial_{i}E_{j}+\partial_{j}E_{i})+2\delta_{ij}\Psi \nonumber\\
&& +\big(\partial_{i}\partial_{j} -\frac{1}{3}\delta_{ij}\partial_{l}\partial^{l}\big)E\bigg],
\end{eqnarray}
it should be noted that all perturbations are functions of time and space, and they agree with the transformations under spatial rotations. Furthermore, we have these conditions $\delta^{ij}h_{ij}=\partial^{i}h_{ij}=\partial^{i}E_{i}=\partial^{i}B_{i}=0$ for scalar, vector, and tensor perturbations which means that the tensor perturbations are transverse and traceless.

We perturb the scalar field $\sigma$ as follows
\begin{equation}
\sigma =\sigma^{(0)}+\delta\sigma.
\end{equation}
It is worth pointing out that the spatial indices on perturbations can be raised and lowered by
$\delta^{ij}$ and $\delta_{ij}$. Also, the actions can be expanded in Fourier plane waves, i.e., $\vec{\nabla}^{2}\rightarrow -k^{2}$, $d^{3}x\rightarrow d^{3}k$.
Note that we show all calculations in the unitary gauge, thus there is not any worry concerning the form of gauge-invariant combinations.

\subsection{Tensor}\label{subTen}

It is noticeable that the tensor perturbations are the only sources of gravitational waves in general relativity. Meanwhile, we know that passing the gravitational waves through spacetime stretches it. The dispersion relation of gravitational waves in modified gravity models changes.
In other words, the propagation speed of the gravitational wave could be different from the speed of light, and the friction term of the tensor perturbations changes too. This way, we calculate the dispersion relation of gravitational waves in the new extension of the dRGT massive gravity theory in the Einstein frame.

We start by considering tensor perturbations around the background,
\begin{equation}
\delta g_{ij}=a^{2}h_{ij},
\end{equation}
where
\begin{equation}
    \partial^{i}h_{ij}=0 \quad {\rm and} \quad g^{ij}h_{ij}=0.
\end{equation}
The tensor perturbed action in the second-order could be obtained for each part of the action separately. The Einstein gravity part of the quadratic perturbed action is,
\begin{eqnarray}
S^{(2)}_{\rm gravity}=\frac{1}{8}\int d^{3}k \, dt \, a^{3}N\Bigg\lbrace \frac{\dot{h}_{ij}\dot{h}^{ij}}{N^{2}}&&-\Big(\frac{k^{2}}{a^{2}}+\frac{4\dot{H}}{N}\nonumber\\&&+6H^{2} \Big)h^{ij}h_{ij}\Bigg\rbrace.\nonumber\\
\end{eqnarray}
The Brans-Dicke part of the perturbed action in quadratic order is
\begin{eqnarray}
S^{(2)}_{\rm Brans-Dicke}=-\frac{1}{8}\int d^{3}k \, dt \, a^{3}N\Bigg\lbrace \frac{(2\omega(\sigma) +3)}{2N^{2}}\dot{\sigma}^{2}h^{ij}h_{ij}\Bigg\rbrace.\nonumber\\
\end{eqnarray}

The massive gravity sector of the perturbed action can be written as
\begin{eqnarray}
S^{(2)}_{\rm massive}&&= \frac{1}{8}\int d^{3}k \, dt \, a^{3}N m_{g}^{2}e^{-2\sigma}\Bigg\lbrace\frac{1}{X_{\rm SA}^{\pm ~ 2}r^{2}N^{2}}\Bigg[(\alpha_{3}+\alpha_{4})rX^{3}\nonumber\\&&
-(1+2\alpha_{3}+\alpha_{4})(1+3r)X^{2}+(3+3\alpha_{3}+\alpha_{4})(3\nonumber\\
&&+2r)X-2(6+4\alpha_{3}+\alpha_{4})\Bigg]\Bigg\rbrace h^{ij}h_{ij}.\nonumber\\
\end{eqnarray}
Summing up the second order pieces of the perturbed actions $S^{(2)}_{\rm gravity}$, $S^{(2)}_{\rm Brans-Dicke}$, and $S^{(2)}_{\rm massive}$, we demonstrate the total action in a second order for tensor perturbations
\begin{eqnarray}
S^{(2)}_{\rm total}=\frac{1}{8}\int d^{3}k \, dt \, a^{3}N\bigg\lbrace \frac{\dot{h}^{ij}\dot{h}_{ij}}{N^{2}}-\Big(\frac{k^{2}}{a^{2}}+M_{\rm GW}^{2}\Big)h^{ij}h_{ij}\bigg\rbrace . \nonumber\\
\end{eqnarray}
Using Eqs. (\ref{XSa}) and (\ref{rS}) we obtained $\alpha_{3}$ and $\alpha_{4}$. Thus, the dispersion relation of gravitational waves is obtained as \\
\begin{eqnarray}
M^{2}_{\rm GW}=\frac{4\dot{H}}{N}+6H^{2}+\frac{(2\omega(\sigma) +3)}{2N^{2}}\dot{\sigma}^{2}+\Xi , \label{eq:M2:GW1}
\end{eqnarray}
where
\begin{widetext}
\begin{eqnarray}
\Xi =\frac{1}{2 r_{SA}^{2}\big[ X_{\rm SA}^{\pm}(X_{\rm SA}^{\pm}(3 r_{SA}-1)-4) +2 \big] (X_{\rm SA}^{\pm}-1)N^{3}} \Bigg\lbrace 2 m_{g}^{2}\bigg[X_{\rm SA}^{\pm ~ 3}\big(3 r_{SA}^{2}-1\big)+6X_{\rm SA}^{\pm ~ 2}\big(1-2 r_{SA}\big)\nonumber\\+6 r_{SA}X_{\rm SA}^{\pm}-2\bigg] N-H^{2} r_{SA}^{2}\bigg(X_{\rm SA}^{\pm}(X_{\rm SA}^{\pm}-3)(r_{SA}X_{\rm SA}^{\pm}-2)-2\bigg)\bigg( N \omega^{'}(\sigma) +6\omega(\sigma)+9\bigg)\Bigg\rbrace.
\end{eqnarray}
\end{widetext}
We demonstrated the modified dispersion relation of gravitational waves.
In fact, the propagation of gravitational perturbations in the FLRW cosmology in the new extension of the dRGT massive gravity in the Einstein frame is presented.

It should be paid attention that if the mass square of gravitational waves is positive, the stability of long-wavelength gravitational waves is guaranteed. But, if it is negative, it must be tachyonic.
Meanwhile, we know that the mass of the tachyon is of the order of the Hubble scale, so, the instability would take the age of the Universe to develop.

Clearly, this result introduces an extra contribution to the phase evolution of gravitational waveform \cite{Will:1997bb,Mirshekari:2011yq}, and can be detected with the accurate matched-filtering techniques in the data analysis. Furthermore, there has been a tendency towards tests of graviton mass after the first discovery of gravitational waves in a merging binary black hole \cite{LIGOScientific:2019fpa,LIGOScientific:2016lio,LIGOScientific:2020tif,Shao:2020shv}. The latest constraint on the graviton mass is around $m_{g}\leqslant 1.76\times 10^{-23} eV/c^{2}$ at 90\% credibility \cite{LIGOScientific:2020tif}. Also, the corresponding Compton wavelength is still much smaller than the Hubble scale, so the relevance to modified cosmology is restricted at present. Using the future space-based gravitational-wave detectors which are much more sensitive to the mass of graviton, we hope that it can be possible to test this essential aspect of gravitation with several gravitational events at different wavelengths \cite{Will:1997bb}.

It is worth mentioning that if we consider only the Einstein-Hilbert part, the dispersion relation of gravitational waves shows the speed of gravitational waves is equal to light.

\subsection{Vector}

In this stage, we would like to perform the vector perturbations analysis in the new extension of the dRGT massive gravity theory in the Einstein frame. \\
It is worth noting that there is evidence that indicates the privileged direction in the Universe. The hemispherical asymmetry and the alignment of the low multi-poles in the CMB are the evidence that shows this issue, and this is the significance of the vector perturbations analysis.
On the other hand, it is obvious that the vector perturbations decay as the Universe expands.
If the initial amplitudes of vector perturbations were so large, these perturbations could have significant amplitudes at present, so, they spoiled the isotropy of the very early Universe. But, in an inflationary universe, we have no large primordial vector perturbations, and they do not have any role in the formation of the large-scale structure of the Universe.
However, the late time's vector perturbations that have been formed after nonlinear structure can explain the rotation of galaxies \cite{Mukhanov:2005sc}.

Here, we consider the vector perturbations,
\begin{eqnarray}\label{Bi}
B_{i}=\frac{a(r^{2}-1)k^{2}}{2\bigg[k^{2}(r-1)+a^{2}(2\omega(\sigma) +3)H^{2}\bigg]}\frac{\dot{E}_{i}}{N}.
\end{eqnarray}
The field $B_{i}$ is a nondynamical, and we can enter it into the action as an auxiliary field.
Therefore, we find a single propagating vector
\begin{eqnarray}\label{AVc}
S_{\rm vector}^{(2)}=\frac{1}{8}\int d^{3}k \, dt \, a^{3}N 
\bigg(\frac{\beta}{N^{2}} |\dot{E}_{i}|^{2} -\frac{k^{2}}{2}M_{\rm GW}^{2}|E_{i}|^{2}\bigg),\nonumber\\
\end{eqnarray}
where
\begin{eqnarray}
\beta =\frac{k^{2}}{2}\bigg(1+\frac{k^{2}(r^{2}-1)}{a^{2}H^{2}(2\omega(\sigma) +3)}\bigg)^{-1}.
\end{eqnarray}
It seems that there are two cases, in the first case, we have $\frac{r^{2}-1}{(2\omega(\sigma)+3)}\geq 0$, and there is no critical momentum scale. 
But in the second case for $\frac{r^{2}-1}{(2\omega(\sigma)+3)}<0$, in order to avoid a ghost, we have a critical momentum scale $k_{c}=\frac{a^{2}H^{2}\big(2\omega(\sigma)+3\big)}{1-r^{2}}$. In other words, to have stability in the system we require the physical critical momentum scale which should be above the ultraviolet cutoff scale of effective field theory, so we have,
\begin{eqnarray}\label{LamC}
\Lambda_{UV}^{2}\lesssim \frac{ H^{2}(2\omega(\sigma) +3)}{1-r^{2}}, \qquad if \quad \frac{(r^{2}-1)}{\omega(\sigma) +\frac{3}{2}}<0.
\end{eqnarray}
In order to determine whether the vector modes suffer from other instabilities, the canonically normalized fields can be considered,
\begin{eqnarray}
\zeta_{i}=\frac{\beta E_{i}}{2}.
\end{eqnarray}
By considering and inserting the above equation in the Eq. (\ref{AVc}), we have
\begin{eqnarray}
S=\frac{1}{2}\int d^{3}k \, dt \, a^{3}N \bigg(\frac{|\dot{\zeta_{i}}|^{2}}{N^{2}}-c_{V}^{2}|\zeta_{i}|^{2}\bigg).
\end{eqnarray}
The sound speed for vector modes is
\begin{eqnarray}\label{c_V}
c_{V}^{2}=M_{GW}^{2}(1+u^{2})-\frac{H^{2}u^{2}(1+4u^{2})}{(1+u^{2})^{2}},
\end{eqnarray}
here we consider the dimensionless quantity as below
\begin{eqnarray}
u^{2}\equiv \frac{k^{2}(r^{2}-1)}{a^{2}H^{2}(2\omega(\sigma) +3)}.
\end{eqnarray}
It is interesting to note that for avoiding tachyonic instability which can be originated from the first part of Eq. (\ref{c_V}), if $M_{GW}^{2}<0$ and $u^{2}>0$, we should consider the below conditions.
\begin{eqnarray}
&&\Lambda_{UV}^{2} \lesssim \frac{H^{2}(2\omega(\sigma) +3)}{r^{2}-1}, \nonumber\\ && if \quad \frac{(r^{2}-1)}{\omega(\sigma) +\frac{3}{2}}>0 \quad and \quad M_{GW}^{2}<0.
\end{eqnarray}
By considering all physical momenta below the UV cut-off $\Lambda_{UV}$, we have a growth rate of instability lower than the cosmological scale.
\\
On the other hand, concerning the second part of Eq. (\ref{c_V}) we have two cases. In the first case, by considering $u^{2}>0$, we do not have instabilities faster than the Hubble expansion. In the second case, if we have $u^{2}<0$, according to the no-ghost condition Eq. (\ref{LamC}), we have $|u^{2}|\lesssim \frac{k^{2}}{a^{2}}\frac{1}{\Lambda_{UV}^{2}}$ to avoid instabilities. Therefore, the second part of Eq. (\ref{c_V}) does not lead to any instabilities.
\\
Finally, it should be pointed out that for avoiding instabilities we should have $c_{V}^{2}>0$ which means that the stability for vector modes is guaranteed. Using this fact, we know that the mass square of the dispersion relation of gravitational waves should be positive as we have mentioned in the Subsection (\ref{subTen}), i.e., ($M_{GW}^{2}>0$).

\subsection{Scalar}

It is interesting to mention that the scalar fields have been introduced to explain the accelerated expansion of the Universe, and they do not break the isotropy of the Universe. The analysis of scalar perturbations contains interesting phenomenology. 
However, in this stage, we want to focus on the stability of the scalar perturbations in the new extension of the dRGT massive gravity theory in the Einstein frame.

We begin with the action quadratic in scalar perturbations
\begin{eqnarray}
\delta g_{00}=&&-2N^{2} \Phi, \nonumber\\
\delta g_{0i}=&&N\,a\,\partial_{i}B, \nonumber\\
\delta g_{ij}=&&a^{2}\bigg[2\delta_{ij}\Psi +\big(\partial_{i}\partial_{j} -\frac{1}{3}\delta_{ij}\partial_{l}\partial^{l}\big)E\bigg],
\end{eqnarray}
\begin{equation}
\sigma =\sigma^{(0)}+\delta\sigma.
\end{equation}
As the perturbations $\Phi$ and $B$ are free of time derivatives, we can eliminate them as auxiliary fields using their equations of motion
\begin{eqnarray}
B=\frac{r^{2}-1}{(3\omega(\sigma) +\frac{9}{2})a H^{2}}\bigg\lbrace H\big[(3\omega(\sigma) +\frac{9}{2})\delta\sigma -6\Phi \big]\nonumber\\+\frac{1}{N}(k^{2}\dot{E}+6\dot{\Psi})\bigg\rbrace,
\end{eqnarray}
\begin{widetext}
\begin{eqnarray}
\Phi =\frac{1}{\bigg[\big(3\omega(\sigma) +\frac{9}{2}\big)\big(\frac{27}{2}-\omega(\sigma) \big) a^{2}H^{2}+12 k^{2}(r^{2}-1)\bigg]} \Bigg\lbrace k^{4}E (\omega(\sigma) +\frac{3}{2})+(3\omega(\sigma) +\frac{9}{2})\bigg(2k^{2}(r^{2}-1)-\frac{(3\omega(\sigma) +\frac{9}{2})a^{2}H^{2}}{r-1}\bigg)\delta\sigma \nonumber\\
+(3\omega(\sigma) +\frac{9}{2})\bigg(2k^{2}+\frac{(3\omega(\sigma) +\frac{9}{2})a^{2}H^{2}}{r-1}\bigg)\Psi - \frac{(3\omega(\sigma) +\frac{9}{2})a^{2}H}{N}\bigg((\omega(\sigma) +\frac{3}{2})\delta\dot{\sigma}-6\dot{\Psi}\bigg)+\frac{2k^{2}}{HN}(r^{2}-1)(k^{2}\dot{E}+6\dot{\Psi})\Bigg\rbrace. \nonumber\\
\end{eqnarray}
\end{widetext}
By substituting these equations into the action, we achieve the action which contains three fields, $E$, $\Psi$ and $\delta\sigma$. Also, we determine another nondynamical combination to remove the sixth degree of freedom, which is
\begin{eqnarray}
\tilde{\Psi}= \frac{1}{\sqrt{2}}(\Psi +\delta\sigma).
\end{eqnarray}
Furthermore, an orthogonal combination can be defined,
\begin{eqnarray}
\tilde{\delta\sigma}=\frac{1}{\sqrt{2}k^{2}}(\Psi -\delta\sigma).
\end{eqnarray}
By redefining these fields, we write the action in terms of $\tilde{\Psi}$, $\tilde{\delta\sigma}$ , and $E$, with no time derivatives on $\tilde{\Psi}$. Therefore, the $\tilde{\Psi}$ is auxiliary and could be eliminated,
\begin{widetext}
\begin{eqnarray}
\tilde{\Psi}=&&\Bigg( -k^{2}-\frac{24 a^{2}H^{2}}{r(r-1)}+\frac{2 a^{2}H^{2}k^{2}\bigg[\big(48-(\omega(\sigma) +\frac{3}{2})(\frac{9}{2}-\omega(\sigma))\big)r-(\omega(\sigma)^{2}+3\omega(\sigma) +\frac{9}{4})\bigg]}{\bigg(4 k^{2}-a^{2}H^{2}(\omega(\sigma) +\frac{3}{2})(\frac{9}{2}-\omega(\sigma))\bigg)(r-1)}\Bigg)\tilde{\delta\sigma}\nonumber\\&&- \frac{2\sqrt{2}k^{4}E}{\bigg[12 k^{2}-3(\omega(\sigma) +\frac{3}{2})(\frac{9}{2}-\omega(\sigma))a^{2}H^{2}\bigg]}+2 a^{2}H\Bigg(\frac{3}{r}+\frac{\bigg(2 k^{2}(r-1)+(3\omega(\sigma) +\frac{9}{2})a^{2}H^{2}\bigg)(\frac{9}{2}-\omega(\sigma))}{\bigg[4 k^{2}-(\omega(\sigma) +\frac{3}{2})(\frac{9}{2}-\omega(\sigma))a^{2}H^{2}\bigg](r-1)}\Bigg)\frac{\dot{\tilde{\delta\sigma}}}{N}\nonumber\\&&+ \frac{k^{2}a^{2}H\sqrt{2}(\frac{9}{2}-\omega(\sigma))}{\bigg[12 k^{2}-3(\omega(\sigma) +\frac{3}{2})(\frac{9}{2}-\omega(\sigma))a^{2}H^{2}\bigg]}\frac{\dot{E}}{N}. \nonumber\\
\end{eqnarray}
\end{widetext}
Note that by substituting this solution in the action and considering the notation $A \equiv (\tilde{\delta\sigma}, E)$, the scalar action can be obtained
\begin{eqnarray}
S = \frac{1}{2}\int d^{3}k \, dt \, a^{3}N\Bigg\lbrace \frac{\dot{A}^{\dagger}}{N}\mathcal{F}\frac{\dot{A}}{N}+\frac{\dot{A}^{\dagger}}{N}\mathcal{D}A \nonumber\\+ A^{\dagger}D^{T}\frac{\dot{A}}{N}-A^{T}\varpi^{2}A\Bigg \rbrace,
\end{eqnarray}
where $D$ is a real anti-symmetric $2\times 2$ matrix, and $\mathcal{F}$ and $\varpi^{2}$ are real symmetric $2 \times 2$ matrices.\\
In the following, we show the components of the matrix $\mathcal{F}$ as below
\begin{widetext}
\begin{eqnarray}
\mathcal{F}_{11}= k^{4}(2\omega(\sigma) + 3)\Bigg[1 +\frac{9 a^{2}H^{2}}{k^{2}(r-1)^{2}} - \frac{a^{2}H^{2}\bigg[ (\omega(\sigma) +\frac{3}{2})+(\frac{9}{2}-\omega(\sigma))r\bigg]^{2}}{\bigg[4 k^{2}-(\omega(\sigma) +\frac{3}{2})(\frac{9}{2}-\omega(\sigma))a^{2}H^{2}\bigg](r-1)^{2}} \Bigg],
\end{eqnarray}
\begin{eqnarray}
\mathcal{F}_{12}= k^{4}\sqrt{2}(\omega(\sigma) + \frac{3}{2})\Bigg[\frac{r}{(\omega(\sigma) +\frac{3}{2})(r-1)}-\frac{k^{2}\bigg[(2\omega(\sigma) +3)+(9-2\omega(\sigma))r\bigg]}{\bigg[12 k^{2}-3(\omega(\sigma) +\frac{3}{2})(\frac{9}{2}-\omega(\sigma))a^{2}H^{2}\bigg](r-1)(\omega(\sigma) +\frac{3}{2})} \Bigg],
\end{eqnarray}
\begin{eqnarray}
\mathcal{F}_{22}= \frac{k^{4}(\omega(\sigma) +\frac{3}{2})}{36} \Bigg[1-\frac{(\frac{9}{2}-\omega(\sigma))^{2}a^{2}H^{2}}{4 k^{2}-(\omega(\sigma) +\frac{3}{2})(\frac{9}{2}-\omega(\sigma))a^{2}H^{2}} \Bigg].
\end{eqnarray}
\end{widetext}
In order to determine the sign of the eigenvalues, we study the determinant of the kinetic matrix $\mathcal{F}$. Thus, we have
\begin{eqnarray}\label{69}
det \, \mathcal{F}\equiv &&\mathcal{F}_{11}\mathcal{F}_{22}-\mathcal{F}_{12}^{2}=\nonumber\\ &&\frac{3 k^{6}(\omega(\sigma)^{2}+3\omega(\sigma) +\frac{9}{4})a^{4}H^{4}}{\bigg[(\omega(\sigma) +\frac{3}{2})a^{2}H^{2}-\frac{4 k^{2}}{(\frac{9}{2}-\omega(\sigma))}\bigg](r-1)^{2}},
\end{eqnarray}
Note that to avoid appearing the ghosts in the scalar sector, we should have
\begin{eqnarray}
\frac{k}{aH} < \frac{\sqrt{(\omega(\sigma) +\frac{3}{2}) (\frac{9}{2}-\omega(\sigma))}}{2}.
\end{eqnarray}
\begin{figure}
\centering
\includegraphics[width=8cm]{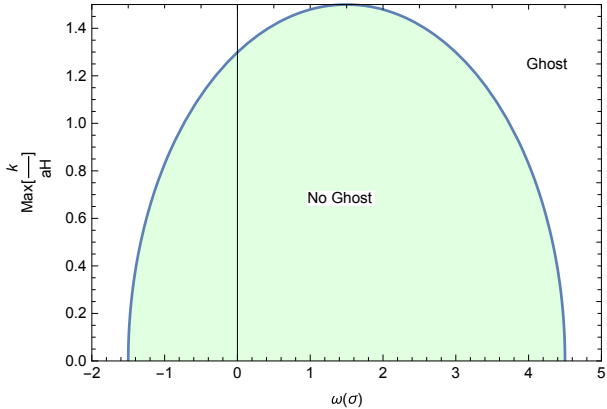}
\caption[figs]
{According to the determinant of kinetic matrix Eq. (\ref{69}), the stability of the scalar sector is imposed. As it can be seen, the below solid line shows there is no ghost degree of freedom which means that the determinant is positive. But, above the solid line, we have a ghost.}
\label{SP}
\end{figure}
As a result, we should note that the stability of the scalar sector is guaranteed using the determinant of the kinetic matrix. In fact, the determinant is positive and we do not have a ghost degree of freedom in the determining part (See figure \ref{SP}).

\section{Conclusion}\label{sec:6}

The significance of this study is the understanding of how the extended theory could be well-behaved and ghost-free in perturbations analysis around their cosmological backgrounds.
In this work, we have introduced the Brans-Dick dRGT massive gravity which is the new extension of massive gravity theory. First of all, we have performed the transformation of the Jordan frame to the Einstein frame, and we have exhibited maintaining the invariance of physical laws under this transformation.

We have presented the total Lagrangian and the full set of equations of motion for an FLRW background. In order to explain the late-time accelerated expansion of the Universe, we have demonstrated the self-accelerating background solution in the context of the new extension of the dRGT massive gravity in the Einstein frame. This way, we have considered the function of $\omega(\sigma)$ instead of a constant to avoid the curvature singularities and a big brake.

Finally, we have analyzed the cosmological perturbations, which consist of tensor, vector, and scalar modes. For studying the mass of graviton for the new extension of the dRGT massive gravity theory, we have calculated the dispersion relation of gravitational waves, and we have shown the propagation of gravitational perturbations in the FLRW cosmology in the Einstein frame. In vector and scalar perturbations, we have presented the conditions of the guaranteed stability of the vector and scalar sectors.

\section*{Acknowledgements}
This paper is published as part of a research project supported by the University of Tabriz Research Affairs Office. We are so thankful for the very nice comments of Professor Gregory Gabadadze. We are also really grateful to Dr. Nishant Agarwal for helpful notes and codes which are related to tensor perturbations. We would like to thank Professor Tina Kahniashvili and Dr. A. Emir Gumrukcuolu for their useful comments.



\bibliography{apssamp}


\end{document}